\documentclass[twocolumn]{aa}

\usepackage{graphicx}
\usepackage{txfonts}
\newcommand{\proton}{\(\mathrm{H^+}\)}
\newcommand{\alphas}{\(\mathrm{He^{2+}}\)}
\newcommand{\water}{\(\mathrm{H_2O^+}\)}
\begin{document}

   \title{Development of a cometosheath at comet 67P/Churyumov-Gerasimenko}
   \subtitle{A case study comparison of Rosetta observations}

   \author{H.~N. Williamson
          \inst{1}
          \and
          H. Nilsson\inst{1}\
          \and G. Stenberg Wieser\inst{1}\ 
          \and A. Moeslinger\inst{1,2}\
          \and C. Goetz \inst{3}\
          }

   \institute{Swedish Institute of Space Physics, Box 812, 981 28 Kiruna, Sweden\\
              \email{hayley.williamson@irf.se}
    \and Department of Physics, Ume\aa\ University, 901 87 Ume\aa, Sweden
    \and European Space Research and Technology Centre, European Space Agency, Keplerlaan 1, 2201AZ Noordwijk, The Netherlands}
   \date{}

 
  \abstract
   {The ionosphere of a comet is known to deflect the solar wind through mass loading, but the interaction is dependent on cometary activity. We investigate the details of this process at comet 67P using the Rosetta Ion Composition Analyzer.}
   {This study aims to compare the interaction of the solar wind and cometary ions during two different time periods in the Rosetta mission.}
   {We compared both the integrated ion moments (density, velocity, and momentum flux) and the velocity distribution functions for two days, four months apart. The velocity distribution functions were projected into a coordinate system dependent on the magnetic field direction and averaged over three hours.}
   {The first case shows highly scattered \proton\ in both ion moments and velocity distribution function. The \alphas\ ions are somewhat scattered, but less so, and appear more like those of \water\ pickup ions. The second case shows characteristic evidence of mass-loading, where the solar wind species are deflected, but the velocity distribution function is not significantly changed.}
   {The distributions of \proton\ in the first case, when compared to \alphas\ and \water\ pickup ions, are indicative of a narrow cometosheath on the scale of the \proton\ gyroradius. Thus, \alphas\ and \water,\ with larger gyroradii, are largely able to pass through this cometosheath. An examination of the momentum flux tensor suggests that all species in the first case have a significant non-gyrotropic momentum flux component that is higher than that of the second mass-loaded case. Mass loading is not a sufficient explanation for the distribution functions and momentum flux tensor in the first case, and so we assume this is evidence of bow shock formation.}

   \keywords{plasmas, 
   methods -- data analysis,
   comets -- individual -- comet 67P/Churyumov-Gerasimenko}

   \maketitle
%

\section{Introduction} \label{sec:intro}

Ion distributions for cometary and solar wind plasmas have been measured at a number of comets, including a flyby of comet 21P/Giacobini-Zinner by the International Cometary Explorer (ICE) mission in 1985 \citep{Brandt1985}, a comet 1P/Halley flyby by the Giotto spacecraft in 1986 \citep{Coates1989,Huddleston1993,Neugebauer1989}, and a subsequent Giotto flyby of comet 26P/Grigg-Skjellerup in 1992 \citep{Coates1993,Grensemann1993}. More recently, the Rosetta mission provided unique, long-term observations of comet 67P/Churyumov-Gerasimenko through perihelion, following it from 2014 to 2016. The size of a cometary ionosphere changes with cometary activity, which varies not only from comet to comet but also within a single comet's orbit. Outgassing at a comet is largely driven by the amount of insolation; thus comet 67P increased in activity with decreasing heliocentric distance \citep{Hansen2016,Nilsson2017}. The long duration of the Rosetta mission therefore presents an opportunity to study the interaction of a comet with the solar wind through varying levels of activity and heliocentric distances, unlike previous flyby missions. 

The Rosetta spacecraft itself primarily stayed within a few hundred km of the comet nucleus. Changing activity levels produced an expanding and subsequently contracting ionosphere. While the spacecraft generally remained at roughly the same distance from the nucleus, the changing levels of comet activity throughout the mission allowed for various solar wind-cometary ionosphere interaction regions to pass over Rosetta \citep{Mandt2016}. Thus, long-term temporal changes in the plasma can be used to identify different portions of the cometary ionosphere-solar wind interaction. This can easily be seen  when comparing the pressures of the different plasma components: cometary ions, solar wind ions, electrons, and magnetic field. For example, from April to December 2015, there is a lack of detectable solar wind ions known as the solar wind ion cavity, even during the 1500 km dayside excursion. Surrounding the solar wind ion cavity, there is a period of roughly four months when the solar wind pressure was equal to or less than that of both the cometary ions and magnetic field. Additionally, during this time period, the solar wind has largely been deflected from its antisunward direction \citep{Behar2017,Williamson2020}. Because of the deflection of the solar wind and balance of the solar wind pressure by the cometary ion and magnetic pressures, this time period is analogous to a magnetopause. Thus, we define the cometopause as the boundary where the flow changes from primarily consisting of the solar wind to primarily consisting of cometary ions, namely where the cometary ion momentum flux becomes greater than the solar wind momentum flux \citep{Coates1997,Gombosi1987,Gringauz1986}. Rosetta crossed the cometopause before and after the solar wind ion cavity, when the solar wind was still observable without dominating the total momentum flux.

Examining the distributions of the cometary ions and solar wind components during this time period can illuminate how the solar wind transfers momentum to the cometary ions. As cometary material is sublimated from the surface, the resulting neutral atmosphere radially expands. A portion of these particles are then ionized via photo- or electron impact ionization, and can then be picked up by the solar wind. In a plasma environment where the gyroradius is smaller than the spatial scale, these pickup ions will gyrate around the interplanetary magnetic field lines because of the Lorentz force and form a ring-beam distribution. The pickup ion distribution is then scattered in pitch angle and energy to form a shell distribution \citep{Huddleston1993,Szego2000}. When waves are present, energy interchange between the ion velocity distribution and waves will create a bispherical shell distribution, with branches centered along the positive and negative phase speeds of the waves -- a distribution that was observed at comet Halley by Giotto \citep{Coates1990a}. The pickup process that creates these distributions includes the transfer of momentum from the high velocity, low-mass solar wind to the low velocity, high-mass pickup ions. The pickup ions thus flow increasingly antisunward with the solar wind, which is deflected around the cometary obstacle in a process known as mass loading. 

The effect of mass loading on the ion velocity distribution functions can become more complicated at a comet, however.  At very low activity, the cometary ionosphere presents no significant obstacle to the solar wind and mass loading has very little effect. At high activity, a situation approaching an induced magnetosphere of a planet occurs. The interaction region begins to behave according to a magnetohydrodynamics (MHD) fluid description, with a bow shock forming, although cometary bow shocks are not straightforward due to the diffuse nature of the cometary obstacle. Bow shocks at comets are a direct consequence of high mass loading and were seen to be weak, with a complex structure, at comets 1P/Halley and 26P/Grigg-Skjellerup. Because of the gradual, slow transition of the solar wind to subsonic speeds at some of these crossings, the broader shock-like structures are also called bow waves \citep{Coates1987,Coates1990,Coates1997}.  Downstream of a bow shock or wave, a region analogous to a magnetosheath at an unmagnetized planet can form. Magnetosheaths are characterized by a transition of the plasma from being dominated by dynamic pressure to being dominated by thermal pressure, as well as a pileup of the magnetic field, leading to a slowing and heating of the solar wind as its gyroradius decreases. Magnetosheaths are regions of unstable and turbulent ion dynamics, causing significant changes to the solar wind \citep{Dubinin2008a,Nilsson2012,Remya2013,Uritsky2011}. The cometosheath, the analogous region at a comet, is the region downstream of the cometary bow shock and it is characterized by a slowing of the solar wind and an increase in plasma density and magnetic field strength \citep{Neugebauer1990a}. 

The type of solar wind-cometary atmosphere interaction is largely determined by the size of the gyroradius compared with the comet-solar wind interaction region. The gyroradius of a newborn water ion in the solar wind magnetic field (a few nT) for a heliocentric distance of 1-2 AU is expected to be on the order of 10,000 km \citep{Nilsson2020}. For a high activity comet like 1P/Halley, with a maximum \(\mathrm{H_2O}\) production rate of \(\sim 2\times 10^{30} \ \mathrm{s^{-1}}\) \citep{Fink1990}, this is well within the scale of the interaction region, as the bow shock was found at a distance of roughly one million km from the nucleus \citep{Coates1987}. However, the gyroradius was generally calculated to be much larger than the interaction region at comet 67P \citep{Nilsson2017}. Previous works have indicated that this resulted in a deflection of the solar wind, but no significant change in speed or energy, as would be expected had a bow shock or cometosheath formed \citep{Behar2017}. However, there have also been indications that a temporary and not fully developed bow shock formed \citep{Gunell2018} that was supported by detections of warm, broadened proton spectra shortly after Rosetta left the solar wind ion cavity (and thus at the point of the mission with the highest cometary activity level where protons could be observed) \citep{Goetz2021}. The observed partial thermalization of the solar wind can be understood as an increase in gyromotion and, hence, a slowing of the bulk flow and transition to thermal energy \citep{Behar2017,Nilsson2017}.  Remote detections of a shock via the electric field \citep{Nilsson2018} and pickup ions distributions \citep{Alho2019,Alho2021} are also possible. Rosetta made a "dayside excursion" during the time the solar wind ion cavity was observed, traveling 1500 km from the nucleus on the dayside, but without observing any solar wind particles. The size of the solar wind ion cavity indicates comet 67P did reach relatively high activity levels during this time. We  would then expect to see heavily mass loaded solar wind and pickup ions just before and after Rosetta was in the solar wind ion cavity, as this will be the highest activity levels for which the solar wind is still observable.

However, even for these higher activity levels, the nominal pickup ion gyroradius (a few thousand km) is still large compared to the interaction region, although the gyroradius decreases when magnetic pileup takes place. This results in complex ion distribution functions, particularly where the flow is the most heavily mass-loaded. At comet 26P/Grigg-Skjellerup, a less active comet than Halley with \(\mathrm{H_2O}\) production rates of \(\sim 7.5\times 10^{27} \ \mathrm{s^{-1}}\) \citep{Johnstone1993} that is still more active than 67P, this resulted in the observation of nongyrotropic particle distributions. At Grigg-Skjellerup, as observed by the Giotto spacecraft, the ion gyroradius was \( \sim 5000 \) km, larger than the length scale of the coma. As a result, the ring distribution one would expect a pickup ion population to form could not be filled completely by the point of observation. These distributions are unstable and likely to excite waves \citep{Motschmann1993}. Several types of non-gyrotropic distributions can result, depending on the specific circumstances, including rotating nongyrotropy, where an incompletely filled ring rotates in phase space at the ion gyrofrequency \citep{Motschmann1997,Motschmann1999}.

The small scale of the interaction region as compared to the gyroradius also affects the ion pressure tensor. For many space plasmas, the pressure tensor can be decomposed into a ``parallel" to the magnetic field component and a ``perpendicular" component. However, for the so-called Finite Larmor Radius situation, a third component becomes important -- the nongyrotropic stress tensor \citep{Stasiewicz1994}. When the pressure tensor is in magnetic field coordinates, this corresponds to the off-diagonal elements, which are directed neither parallel nor perpendicular to the magnetic field. Much like nongyrotropic particle distributions, this is expected to occur at a comet. Nongyrotropic pressures have also been seen in the magnetosheaths of planets such as Mercury and Venus \citep{Phillips1987,Uritsky2011}.

In this study, we compare the distribution functions and momentum flux tensors (i.e., the sum of the dynamic and thermal pressures) of two days during the Rosetta mission. The first, on January 23, 2016, was approximately one month after the end of the solar wind ion cavity and so, the spacecraft was located inside the cometopause. The second, May 10, 2016, was during a lower activity phase of the comet, well past its perihelion. Thus, the first case shows a more heavily mass-loaded and disturbed plasma, while the second exhibits the early stages of the pickup process. Section \ref{sec:methods} introduces the data used and our methods. Section \ref{sec:results} shows the results of our analysis and Section \ref{sec:discussion} presents a discussion. Section \ref{sec:conclusions} concludes the paper. 

\section{Methods} \label{sec:methods}
\subsection{Data} \label{sec:data}
This work primarily uses data from the Rosetta Plasma Consortium (RPC) Ion Composition Analyzer (ICA), an ion mass spectrometer designed to measure the three-dimensional velocity distribution functions of positive ions and capable of distinguishing between \proton, \alphas, \(\mathrm{He^+}\), and heavy ions, primarily \water\ products. Throughout this paper, we use \water\ as shorthand for heavy ions near mass 18 originating from the comet. However, this can also include species such as \(\mathrm{H_3O^+}\) and \(\mathrm{NH_4^+}\), with \water\ generally being the dominant species \citep{Beth2020}. The dynamics remain the same for all of these ions, as it is dependent on mass per charge, and so \water\ is used for simplicity. The ICA sensor measures a nominal energy range of a few eV to 40 keV with an energy resolution of 7\%. It has 16 \( 22.5^{\circ} \times 5.6^{\circ} \) azimuthal anode sectors and 16 elevation angles up to \(\pm 45^{\circ}\), giving a total of \(360^{\circ}\times 90^{\circ}\) field of view. However, due to spacecraft shadowing, the actual total field of view is roughly \(2 \pi \) sr. One full scan of all energies, azimuthal sectors, and elevation angles takes 192 s \citep{Nilsson2007}. In this study, we additionally distinguish between \water\ pickup and newborn ions by dividing the heavy mass bins at 60 eV, an approximate minimum energy for the pickup ions \citep{Bercic2018,Nilsson2020}. We use the mass-separated differential flux data as input for both our momentum flux and distribution function calculations.

We additionally used data from the Rosetta magnetometer (MAG), which consists of two triaxial fluxgate magnetometers on a 1.5 m boom. The maximum sample rate and resolution are 20 vectors per second and 31 pT, respectively, with uncertainties of a few nT per component due to spacecraft magnetic noise \citep{Glassmeier2007,Goetz2016,Richter2019}.

To ensure that the plasma is within the ICA field of view, we also compared the ICA data to data from the RPC Ion and Electron Spectrometer (IES), which consists of two analyzers, one for ions and one for electrons, with an energy range of 4 eV/q to 17 keV/q and energy resolution of 8\%. In addition, IES does not separate particles by mass. Similarly to ICA, IES has a field of view of \(360^{\circ}\times 90^{\circ}\) \citep{Burch2007}. However, because IES was positioned differently on the spacecraft, the field of view of IES is tilted with respect to the ICA field of view by \(60^{\circ}\). Thus, the IES field of view covers some areas that ICA cannot and vice versa. Work is ongoing to combine these two datasets for a fuller understanding of the plasma; for the current study, we  used IES data to confirm that the plasma is within the ICA field of view. The two instruments combined have a field of view of approximately 3.5\(\pi\) sr; thus, there is a high likelihood that any particle distribution will be observed by one or both of the instruments. To confirm the validity of the data used in this study, the IES measurement data were rotated into the same coordinate system as the ICA data. This shared reference frame was used to identify the main directions of the different ion populations and verify that no significant ion population was missed by ICA. As IES had lower angular resolution (\(12^{\circ} \times 45^{\circ}\) per pixel for the data in this study) than ICA, it was not used for calculation of the velocity distribution function or momentum flux and it is not shown in this paper. 

\subsection{Velocity distribution function}
The ICA data is used in two ways to compare our case studies. Firstly, the velocity distribution function is calculated from the observed differential flux:
\begin{equation} \label{eq:fv}
f(\vec{r}, \vec{v}) = \frac{m^2}{2E} \ j(E, \Omega) 
\end{equation}
where \(j\) is the differential flux derived from measured count rates. One value will be obtained for each of the 96 energy bins, 16 sector bins, and 16 azimuth bins. For this study, the instrument spherical coordinate system \((E, v_{\phi}, v_{\theta})\) was converted to Cartesian coordinates \((v_x, v_y, v_z)\). This is done by creating a three dimensional grid and mapping each instrument coordinate bin to the appropriate Cartesian grid pixel \citep{Behar2017}. The mapping is repeated separately for each scan and species. We chose to use a \(50 \times 50 \times 50 \) pixel grid, which gives sufficient resolution that each instrument pixel maps to a separate Cartesian pixel. The endpoints of the grid are chosen separately for each species, as they have varying maximum and minimum velocities. For \proton\ and \alphas, the maximum and minimum values are \(\pm 600\) km/s, while for \water\ it is \(\pm 60\) km/s.

The Cartesian grid is rotated so that the \(v_x\) axis is pointed sunward, and the \(v_y\) axis is along the magnetic field component perpendicular to \(v_x\). Because the undisturbed solar wind velocity is in the antisunward (\(-v_x\)) direction, the \(v_z\) axis orthogonal to \(v_x\) and \(v_y\) will be along the approximate convective electric field due to \( \vec{E} = -\vec{v} \times \vec{B} \), defined as the comet-sun-electric (CSE) coordinate system. Defining CSE coordinates using the upstream, undeflected solar wind velocity such that the x-axis points sunward is in line with previous works (see e.g. \citet{Behar2017,Edberg2019,Goetz2021,Masunaga2019}, etc.). While velocity distribution functions are frequently depicted in a two-dimensional reference frame parallel and perpendicular to the magnetic field, here we chose to preserve the electric field and sunward directions separately. The particle velocities and magnetic field in Section \ref{sec:results} are shown in cometocentric solar equatorial (CSEQ) coordinates, where x points sunward, z is ecliptic north, and y completes the right-handed system. 

\subsection{Momentum flux} \label{sec:momflux}
We additionally use the ion moment data, where the zeroth order is density, first order is velocity, and second order is the pressure tensor:
\begin{equation}\label{eq:moment}
\mathbf{M}^k = \int_{\vec{v}} \, \vec{v}^k \ f(\vec{v}) \ d^3 \vec{v} 
\end{equation}
where \(v = \sqrt{2 E / m}\) and \(f(\vec{v})\) is the distribution function from Eq. \ref{eq:fv}. 

If the first order bulk velocity is \(\vec{V}\), then the second order moment pressure is:
\begin{equation} \label{eq:pressure} 
\tens{P} = m \int_{\vec{v}} (\vec{v} - \vec{V})(\vec{v} - \vec{V}) \, f(\vec{v}) \ d^3 \vec{v}
\end{equation}
where each tensor element can be simplified to
\begin{equation} \tens{P}_{ij} = m \, \tens{M}^2_{ij} - nm V_i V_j \end{equation}
where
\begin{equation} \label{eq:momflux}
\tens{\Phi}_{Pij} \equiv m \tens{M}^2_{ij} = m \int_\Omega \int_{\vec{v}} v_i v_j \, f(\vec{v}) \ d\Omega \, d \vec{v} 
\end{equation} 
is the momentum flux, \(i, j \in \{x, y, z\} \) iteratively, \(n\) is the species number density, and \(m\) is the ion mass (see \citet{BeharThesis}, \citet{Franz2006}, \citet{Nilsson2020} Appendix A, and \citet{Paschmann1998}for full moment calculations). We used Eq. \ref{eq:momflux} in our calculations, as the plasma bulk properties are not species-independent, but must be calculated from the total plasma and weighted by species mass \citep{Gunell2017a}. Due to the higher mass and density of \water, the bulk properties will then be largely dependent on the cold cometary ions. While ICA can theoretically measure down to very low energies, in practice these low energy ions are distorted by the spacecraft potential, making their distributions difficult to interpret \citep{Bergman2020,Bergman2020a}. This is particularly true for a high negative spacecraft potential, which was common throughout the mission \citep{Odelstad2017,Johansson2020}. Also, ICA has seen considerably lower total ion densities than the Langmuir Probe and Mutual Impedance Probe \citep{Gunell2017a}. For the time periods in this work, the LAP/MIP electron densities are nearly two orders of magnitude greater than the ion densities seen by ICA (see Figs. \ref{fig:vel1}(d) and \ref{fig:vel2}(d)). An analysis of cold ion detections strongly suggest that part of the cold ion population subsequently falls outside of the ICA FOV and so it is assumed that ICA is not detecting a significant portion of the cold plasma population \citep{Bergman2021}. Because of this, we cannot calculate a properly weighted bulk velocity and guiding center of mass, meaning any pressure calculation via Eq. \ref{eq:pressure} would be incomplete. Momentum flux, however, does not require the subtraction of the bulk dynamic pressure, and is thus possible to calculate. The momentum flux provides similar information; namely, where the momentum and energy of the plasma is directed and when it is transferred between species.

The momentum flux tensor as given in Eq. \ref{eq:momflux} then has the form:
\begin{equation}
 \tens{\Phi_P}  = \begin{bmatrix}
    \Phi_{Pxx} & \Phi_{Pxy} & \Phi_{Pxz} \\
    \Phi_{Pyx} & \Phi_{Pyy} & \Phi_{Pyz} \\
    \Phi_{Pzx} & \Phi_{Pzy} & \Phi_{Pzz} 
    \end{bmatrix} 
\end{equation}
where the tensor is symmetric such that \( \Phi_{Pij} = \Phi_{Pji} \). The tensor for each scan is then rotated from ICA instrument coordinates to CSE coordinates as described above, such that \(\hat{y} = \hat{B}\) and \(\hat{z} = \hat{E}\). For a typical magnetized space plasma, the off-diagonal elements of this tensor are negligible. However, because the gyroradii at comet 67P are large compared to the spatial scale of the interaction region, the off-diagonal elements are frequently close in magnitude to that of the diagonal elements. This nongyrotropic off-diagonal component is also called the Finite Larmor Radius correction for this reason \citep{Stasiewicz1994}. In Section \ref{sec:results} we include a comparison of the magnitudes of the diagonal and off-diagonal components of the tensor, including both the upper and lower off-diagonal elements, in CSE coordinates:
\begin{equation} 
\frac{|\Phi_{Pdiag}|}{|\Phi_{Poff}|} = \frac{\sqrt{\Phi_{Pxx}^2 + \Phi_{PBB}^2 + \Phi_{PEE}^2}}{\sqrt{\Phi_{PxB}^2 + \Phi_{PxE}^2 + \Phi_{PBx}^2 + \Phi_{PBE}^2 + \Phi_{PEx}^2 + \Phi_{PEB}^2}} 
\end{equation}

\section{Results} \label{sec:results}
\begin{figure}
    \centering
    \includegraphics[width=\hsize]{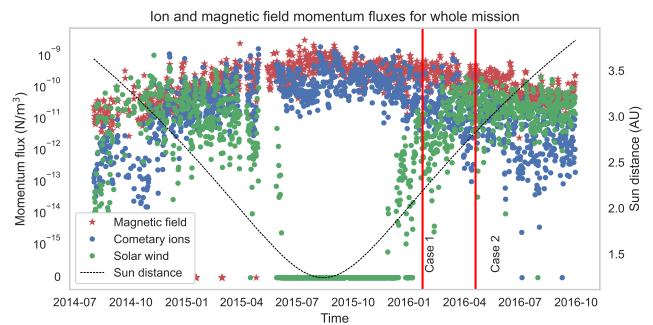}
    \caption{Momentum flux for the magnetic field, cometary ions (\water\ group ions) of all energies and solar wind ions (\proton, \alphas, and \(\mathrm{He^+}\)) for the entirety of the Rosetta escort phase. Each point is the average of 12 hours of data (the rotation period of the comet) to eliminate any effects due to asymmetrical outgassing. The dashed line indicates the heliocentric distance of the comet. The two case study dates are marked by vertical red lines.}
    \label{fig:whole_mission}
\end{figure}

For this study, we chose two dates from the escort phase of the Rosetta mission, January 23, 2016 (case 1) and May 10, 2016 (case 2). These two dates were chosen based on the calculated momentum flux of the cometary and solar wind ions as described in \citet{Williamson2020}. The momentum flux magnitude of these two plasma populations changes with heliocentric distance due to changing cometary activity, as shown in Fig. \ref{fig:whole_mission}, which gives the momentum flux of the solar wind species and cometary ion species averaged over 12 hours, which is approximately the rotation period of the comet. The solar wind ion cavity is clearly visible as the time period when the solar wind momentum flux magnitude drops to zero. When cometary activity is low, the solar wind momentum flux is dominant and the magnetic pressure (\(P_B = \tfrac{B^2}{2 \mu_0}\)) is correlated with the solar wind \citep{Williamson2020}. However, just before and after the solar wind ion cavity, there are time periods when the solar wind is present, but has lower momentum flux than the cometary ions. The magnetic pressure is also greater than the solar wind momentum flux, indicating an increase in the magnetic field. The spacecraft is thus observing a mass loaded ionosphere, where the solar wind has transferred much of its momentum to the pickup ions. 

In order to investigate the solar wind-cometary atmosphere interaction in more detail, we select one day from within the cometopause that exhibited an unusually widespread \proton\ signal in the full day energy-sector spectra and a second day at random from dates when cometary activity had lowered sufficiently such that the momentum flux is dominated by the solar wind. Both are marked in Fig. \ref{fig:whole_mission} with vertical red lines, showing that the average solar wind momentum flux is lower than that of the cometary ions for case 1 and vice versa for case 2. The distance from the sun, gas production rate from the Rosetta Cometary Pressure Sensor (COPS), and mean magnetic field strength for both dates are shown in Table \ref{tab:cases}. Case 1 has a gas production rate of an order of magnitude greater than that of case 2, four months later. Additionally, the magnetic field strength in case 1 is nearly twice that of case 2 for case 1 due to magnetic field pileup. The spacecraft was on the dawn side of the comet with a solar zenith angle of \(\sim 50^{\circ}\) for case 1 and in the dusk terminator plane for case 2. The spacecraft was much farther from the comet in case 1 than case 2 due to the higher activity levels and thus more extended atmosphere. We note that the higher production rate for case 1 is still roughly an order of magnitude smaller than that of Grigg-Skjellerup and four orders of magnitude smaller than the production rate of comet Halley, as observed by Giotto.

\begin{table}[ht]
\caption{Dates, heliocentric distance, mean radial distance of the spacecraft from the nucleus, and comet gas production rate for both cases analyzed in this paper.}
\label{tab:cases}
\centering
\begin{tabular}{lrrrrr}
\hline
          & Date  & $R_{S}$ (AU) & $|r_{sc}|$ (km) & Q (\#/s)  \\
\hline
Case 1 &    2016-01-23  & 2.2  & 76.7    & 5.8e+26             \\
Case 2 &    2016-05-10  & 3.0  & 13.3    & 6.5e+25             \\
\end{tabular}
\end{table}

\subsection{Ion moments}

\begin{figure}[ht]
    \centering
    \includegraphics[width=\hsize]{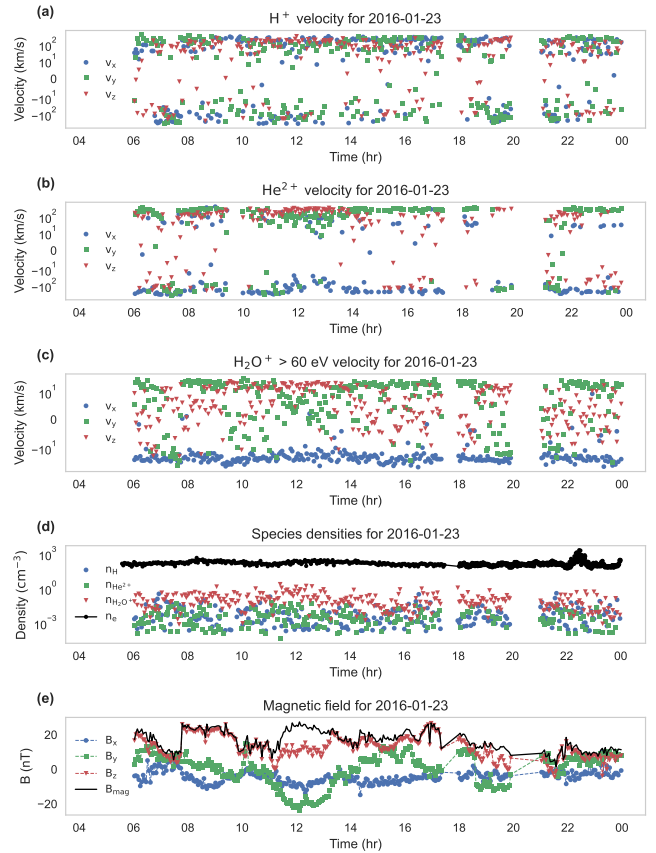}
    \caption{Velocities of \proton, \alphas, \water > 60 eV, densities for each plus the density from the LAP/MIP instruments, and the magnetic field for the date January 23, 2016. All vectors are shown in CSEQ coordinates, where x (blue) is sunward, z (red) is ecliptic north, and y (green) is orthogonal to both. The solid black line in panel (e) shows the magnitude of the magnetic field.}
    \label{fig:vel1}
\end{figure}

\begin{figure}[ht]
    \centering
    \includegraphics[width=\hsize]{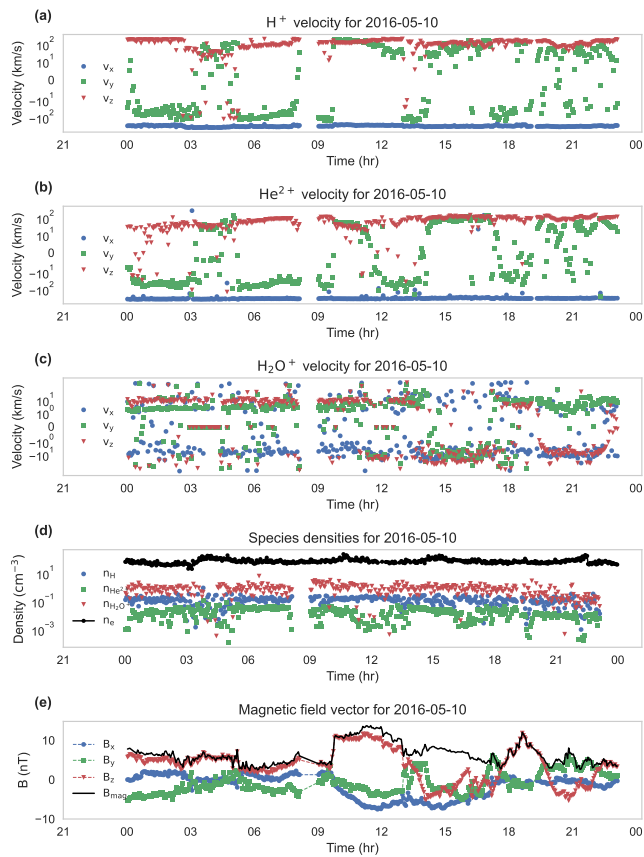}
    \caption{Velocities of \proton, \alphas, \water, densities for each plus the density from the LAP/MIP instruments, and the magnetic field for the date May 10, 2016. Panels (c) (d) shows velocities and density respectively for \water\ of all energies. All vectors are shown in CSEQ coordinates, where x (blue) is sunward, z (red) is ecliptic north, and y (green) is orthogonal to both. The solid black line in panel (e) shows the magnitude of the magnetic field.}
    \label{fig:vel2}
\end{figure}

We first examined the flow direction for the species of interest in our cases: \proton, \alphas, and \water\ in Figs. \ref{fig:vel1} and \ref{fig:vel2}. While ICA also detected \(\mathrm{He^+}\), densities were generally significantly lower than that of the other two solar wind species, as it is the result of charge exchange, and so this is not of interest for the study of mass loading. Figure \ref{fig:vel1} shows only \water\ > 60 eV. The choice of energy cutoff for \water\ is to ensure we are studying the pickup ion population \citep{Bercic2018}. Additionally, as described in section \ref{sec:methods}, ICA does not always capture the full distribution for low energy \water, and so data for low energies are unlikely to provide the full picture. However, for case 2, the \water\ energies were generally lower than in case 1, and there are few detections of \water > 60 eV. Therefore, we show the velocities of all \water\ in Fig. \ref{fig:vel2}. The majority (\(\sim\)85\%) of the \water\ ion energies exceed those that would be affected by the spacecraft potential, so the overall velocity directions should not be distorted as described in Sec. \ref{sec:methods}. All velocities and magnetic field are plotted in CSEQ coordinates. There is a gap in the data for case 1 until 06:00 UST, so the full 24 hours are not shown for January 23.

There are a few major differences visible between Fig. \ref{fig:vel1} and Fig. \ref{fig:vel2}. In panels (a) and (b) of Fig. \ref{fig:vel2}, the velocity of the solar wind looks as we would expect -- dominated by a -x (antisunward) component, in blue dots. While there are significant components along the y and z directions, indicating that the solar wind is being deflected by the cometary obstacle, it is generally nearly an order of magnitude lower than the x component. This is in contrast to panels (a) and (b) of Fig. \ref{fig:vel1}, where all the directional components of the solar wind vary wildly. One feature that is especially noteworthy is (particularly for \proton\ in Fig. \ref{fig:vel1}(a)) that the x component is more often positive than negative, indicating a frequently sunward-directed solar wind. This can be compared to \water\ in Fig. \ref{fig:vel1}(c), where the x component is nearly always antisunward and larger than the y and z components, much like the solar wind from case 2 in Fig. \ref{fig:vel2}(a) and (b), although the speed is much lower for \water. By including \water\ of all energies in Fig. \ref{fig:vel2}(c), we also include those ions which are still radially expanding from the nucleus, meaning velocities are similar in all directions. The flow directions in Fig. \ref{fig:vel1} support that case 1 is a highly disturbed, mass-loaded flow, where the cometary ions now have a significant antisunward component like that of the nominal solar wind. The solar wind in this case, however, has been significantly deflected. Meanwhile, case 2 in Fig. \ref{fig:vel2} shows an antisunward flow direction for the solar wind, with the cometary ions moving in more scattered directions and with more significant y and z components.

\subsection{Ion distributions}
Because changes in magnetic field will affect the ion velocity distribution functions, we select a portion of each day with a mostly unchanging \(\vec{B}\) for our study of the ion distributions: 10:00 - 14:00 UST for case 1 and 00:00 - 03:00 UST for case 2, 56 scans of 192 s each in total for both cases. With a time period of relatively stable magnetic field, any changes in the velocity distribution function should be due to plasma properties rather than a rotation of the magnetic field. We show the distributions plotted in CSE coordinates as described in Section \ref{sec:methods}.

\begin{figure*}[ht!]
\centering
    \resizebox{0.85\hsize}{!}
        {\includegraphics{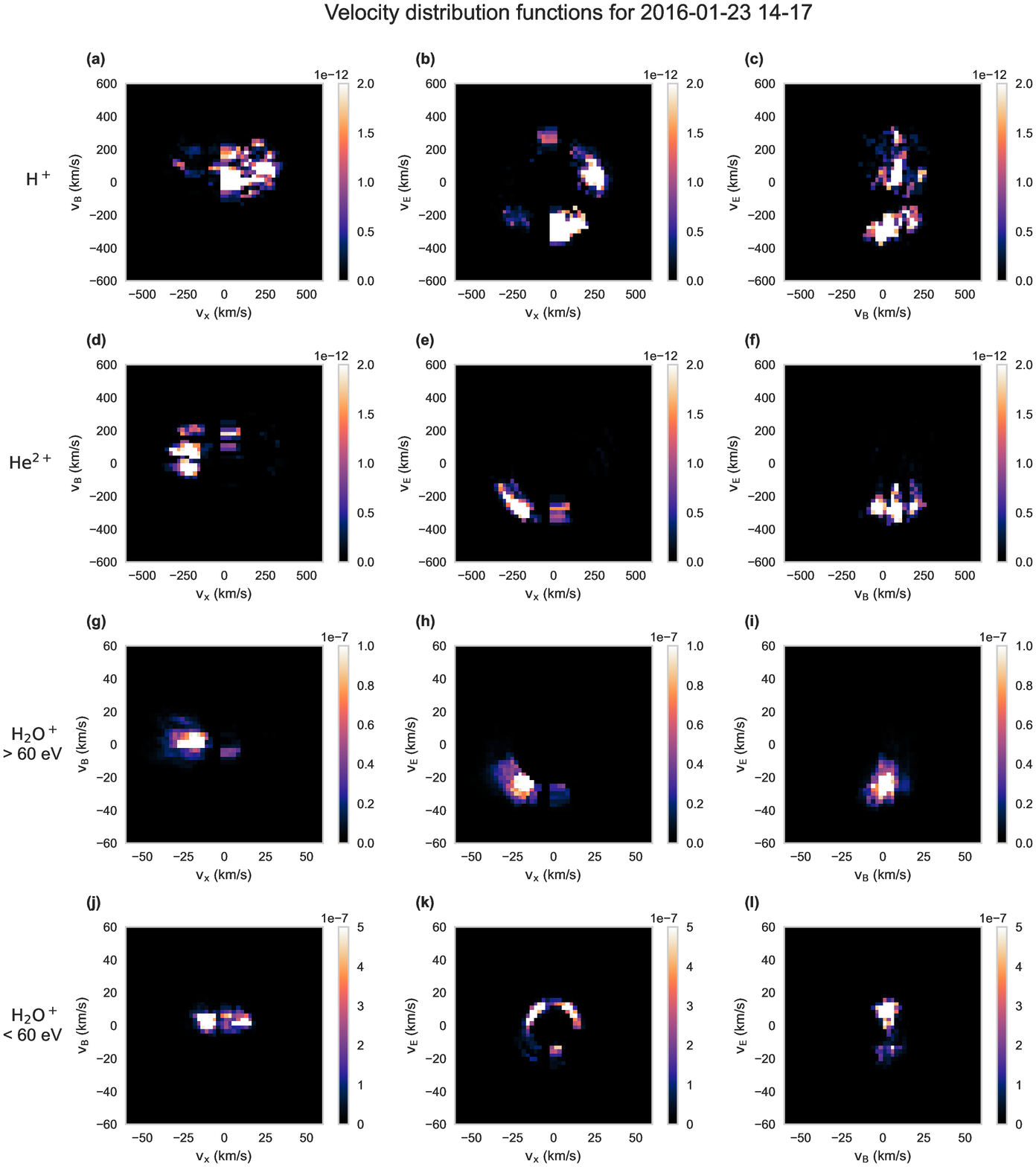}}
        \caption{Distribution functions for three species as described in the text, with \water\ split by energy. The first row is \proton\ (a-c), the second \alphas\ (d-f), third \water > 60 eV (g-i), and fourth \water < 60 eV (j-l). Each column is a projection onto a different pair of axes, with the first column the \(v_x-v_B\) plane, second \(v_x-v_E\) plane, and third \(v_B-v_E\) plane. The color indicates value of the distribution function, bin-averaged over 3 hours (14:00 - 17:00) on 2016-01-23. Color bars are in units of \(\mathrm{s^3/m^6}\).}
     \label{fig:case1}
\end{figure*}

\begin{figure*}[ht!]
\centering
    \resizebox{0.85\hsize}{!}
    {\includegraphics{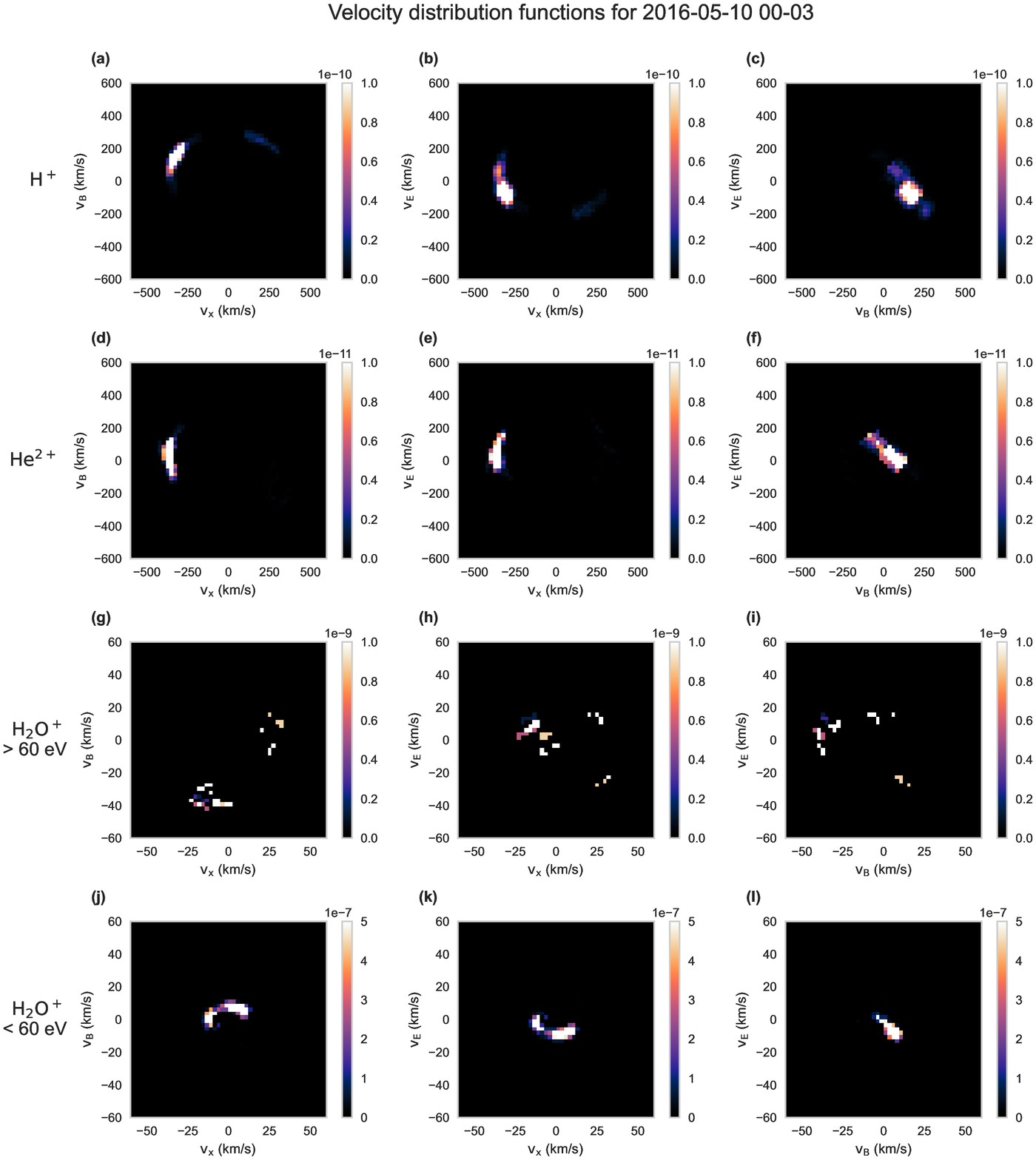}}
    \caption{Distribution functions for three species as described in the text, with \water\ split by energy. The first row is \proton\ (a-c), the second \alphas\ (d-f), third \water > 60 eV (g-i), and fourth \water < 60 eV (j-l). Each column is a projection onto a different pair of axes, with the first column the \(v_x-v_B\) plane, second \(v_x-v_E\) plane, and third \(v_B-v_E\) plane. Colors indicate values of the distribution function, bin-averaged over 3 hours (00:00 - 03:00) on 2016-05-10. Color bars are in units of \(\mathrm{s^3/m^6}\).}
    \label{fig:case2}
\end{figure*}

The plots of the distribution functions in Figs. \ref{fig:case1} and \ref{fig:case2} are the results of multiple ICA scans summed together over a time period of seven hours for both cases. While this gives a sense for the general distribution of the plasma in the region, since individual scans often contain few or no data points, it obscures more time-dependent features, which change from scan to scan. Therefore, we include videos where each frame is a single ICA scan for panels (b), (e), and (h) in both Figs. \ref{fig:case1} and \ref{fig:case2}. 

Beginning with \proton, panels (a), (b), and (c) of Fig. \ref{fig:case1} show \proton\ that have been highly scattered in both phase angle and energy. The \proton\ move both along and against the electric and magnetic fields. This distribution is highly unusual for solar wind \proton\ detected by ICA, such as those shown in panels (a), (b), and (c) of Fig. \ref{fig:case2}. For the second case, the \proton\ approach a narrow beam-like distribution primarily antisunward (\(-v_x\)), with a slight deflection in the \(-v_E\) direction. This distribution is more typical for the solar wind, and so the one in Fig. \ref{fig:case1} stands out. 

The \alphas\ in panels (d), (e), and (f) of Fig. \ref{fig:case1} show slightly less scattering than \proton. In particular, they largely remain in the \(-v_x\) and \(-v_E\) directions, although they subtend a broader phase angle than shown in the corresponding panels of Fig. \ref{fig:case2}, where \alphas\ are even more beam-like than \proton. For both species, the maximum phase space density is in general roughly two orders of magnitude lower for case 1 in Fig. \ref{fig:case1} than in Fig. \ref{fig:case2}, due to the lower fluxes of the solar wind species for the first case.

The \water\ pickup ions in case 1 (Fig. \ref{fig:case1} panels (g), (h), and (i)), on the other hand, are less scattered than either solar wind species, and, in fact, look most similar to \proton\ in Fig. \ref{fig:case2} from case 2, albeit with velocities approximately an order of magnitude lower. They are primarily moving in the antisunward \(-v_x\) direction with some deflection in \(-v_E\). For the second case in Fig. \ref{fig:case2}(g), (h), and (i) there are few detections of \water\ above 60 eV as mentioned previously. Thus the distributions are sparse and difficult to interpret. However, they appear to have been scattered widely in both angle and energy into partial shell distributions, frequently directed in the \(+v_E\) direction along the convective electric field. 

The low energy \water\ in panels (j), (k), and (l) of both Figs. \ref{fig:case1} and \ref{fig:case2} are most likely a type of cold Maxwellian distribution, with the appearance of a ring due to acceleration of the lowest energy ions by the spacecraft potential and subsequent distortion of the field of view. Likewise, while both low energy \water\ distributions have a slight asymmetry, it cannot be ruled out that this is also due to field of view distortions. Comparisons of ICA data to LAP and MIP data suggest that Maxwellian distributions are appropriate for this low energy population \citep{Gunell2017,Gunell2017a}. The mean spacecraft potential is -11.4 V for case 1 and -7.8 V for case 2, corresponding to velocities of \(\sim 11\) km/s and \(\sim 9\) km/s, respectively. Thus, the gap in the center of the low-energy cometary ion distributions is due to the acceleration of the low energy ions by the spacecraft potential, rather than a real feature.

\begin{figure*}[ht]
    \centering
    \includegraphics[width=0.85\hsize]{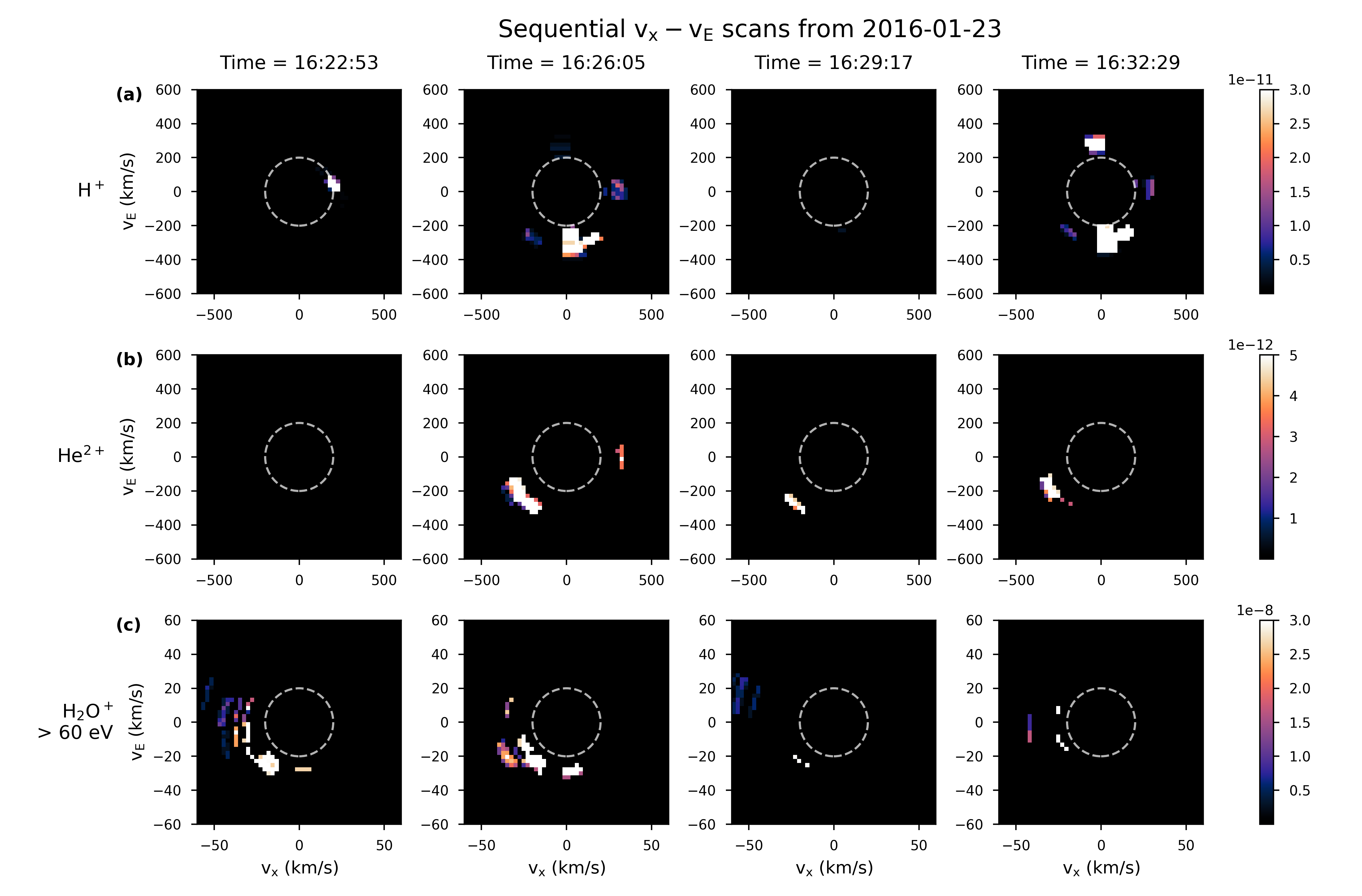}
    \caption{Sequence of four ICA scans of \proton\ (row (a)), \alphas\ (row (b)), and \water > 60 eV (row (c)) on January 23, 2016 projected on the \(v_x-v_E\) plane. Color indicates value of the distribution function and the color bar is in units of \(\mathrm{s^3/m^6}\). The dashed circle indicates a velocity of 200 km/s in rows (a) and (b) for \proton\ and \alphas\ and 20 km/s in row (c) for \water.}
    \label{fig:seq_case1}
\end{figure*}

While the \proton\ and \alphas\ in Fig. \ref{fig:case1} superficially appear to be shell distributions, this is a sum over seven hours and is not necessarily representative of the instantaneous distribution. The online movies show the scan-by-scan distributions for the \proton, \alphas, and \water > 60 eV in the \(v_x-v_E\) plane, where each frame is a single ICA scan. These indicate that, particularly for \proton, the distribution is rotating in time. Four sequential frames from the movie that illustrate this for \proton\ in the \(v_x-v_E\) plane are shown in Fig. \ref{fig:seq_case1}. The sequential frames include a dashed circle at 200 km/s for \proton\ and \alphas\ and 20 km/s for \water\ to illustrate that, in addition to rotating and having a spread in phase angle, there are also changes in energy. Figure \ref{fig:seq_case1} shows how drastically the distribution of \proton\ can change between ICA scans. While there is also some movement for \alphas, the phase angle and energy change are much less. The \water\ pickup ions show a spread in energy, as expected, and the amount of change in phase angle is comparable to \alphas.

\begin{figure}
    \centering
    \includegraphics[width=\hsize]{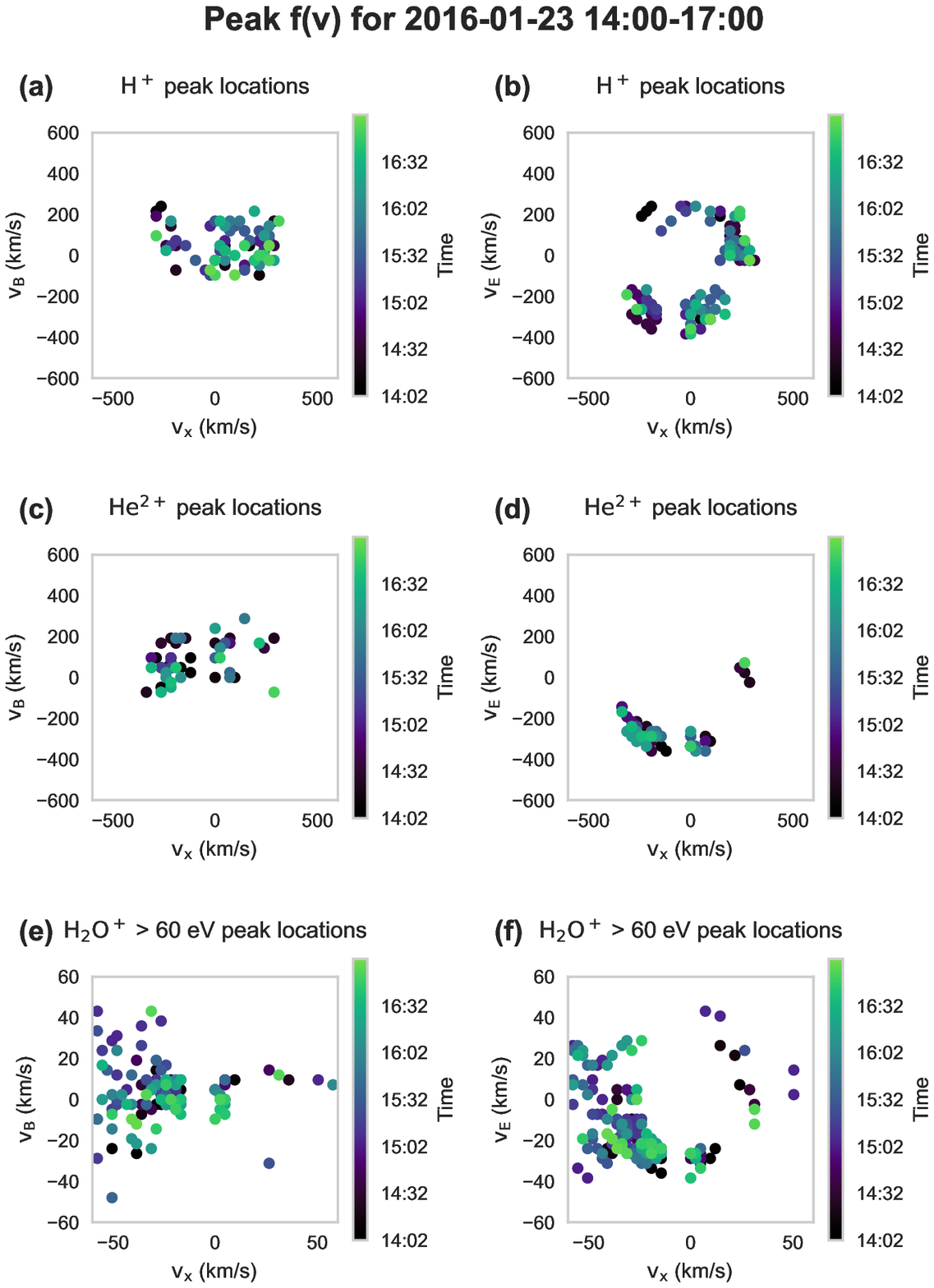}
    \caption{Locations of the pixels with maximum values for each ICA scan in case 1. Top-row, panels (a) and (b) show the \proton, the second row \alphas, and the third row \water > 60 eV. The left column is the \( v_x-v_B\) plane and the right column the \(v_x-v_E\) plane. Color indicates time of the scan.}
    \label{fig:case1_peaks}
\end{figure}

\begin{figure}
    \centering
    \includegraphics[width=\hsize]{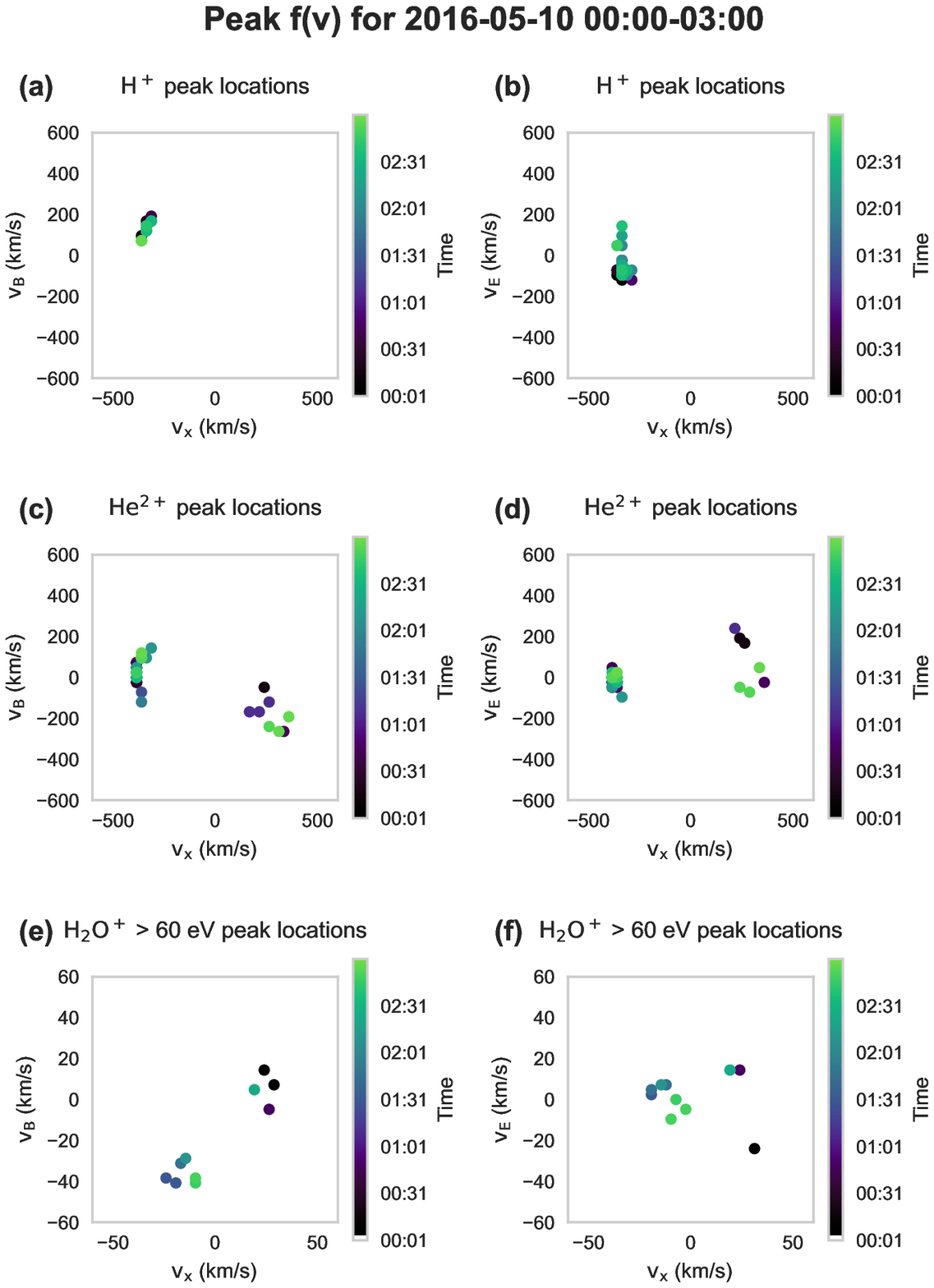}
    \caption{Locations of the pixels with maximum values for each ICA scan in case 2. Top-row, panels (a) and (b) show the \proton, the second row \alphas, and the third row \water > 60 eV. The left column is the \( v_x-v_B\) plane and the right column the \(v_x-v_E\) plane. Color indicates time of the scan.}
    \label{fig:case2_peaks}
\end{figure}

To further examine the time evolution of the distribution in each scan, we plotted the location of the local maxima of the distribution in Figs. \ref{fig:case1_peaks} and \ref{fig:case2_peaks}. Our algorithm picks the local pixel with the highest distribution function value, then returns the coordinates of that pixel if it is within an order of magnitude of the global maximum. This threshold was chosen to account for some scans having multiple peaks, such as those seen in frames 2 and 4 of Fig. \ref{fig:seq_case1} row (a) for \proton.The highest local pixel locations are then plotted for all the ICA scans, so that the movement of the peak of the particle distribution is visible. Figure \ref{fig:case1_peaks} shows the projection on the \(v_x-v_B\) and \(v_x-v_E\) planes for case 1 to show movement relative to magnetic and electric field directions, while Fig. \ref{fig:case2_peaks} shows the projections for case 2. For most scans, there is a single peak; however, some scans (such as for \alphas\ in panels (c) and (d) for in Fig. \ref{fig:case2_peaks}) have two peaks located some distance apart. However, as can be seen when comparing Fig. \ref{fig:case2_peaks}(c) and (d) to Fig. \ref{fig:case2}(d) and (e), these secondary peaks in the \(+v_x\) direction are generally much smaller in value than the primary peak in the \(-v_x\) direction and only appear in a few scans. Because of the smaller magnitude, they are not visible in the averaged plots.

In Fig. \ref{fig:case1_peaks}(a) and (b), the projections along \(v_B\) and \(v_E\), respectively, of \proton, shows that this is not a shell distribution as it may appear, but rather one that moves in time. The distribution peak appears to be rotating around both the magnetic and electric field directions. The gyroperiod of all species for both cases is a few seconds, much smaller than the ICA scan time of 192 s, so it is not possible to obtain a true rotation frequency from the data. However, because we have selected a time period with a stable, non-rotating magnetic field, it is also unlikely this rotation is caused by changes in the magnetic field. As will be discussed in section \ref{sec:discussion}, this is likely due to observing \proton\ in different stages of gyration with a low drift velocity, evidenced by the distribution being roughly centered at 0 km/s.

The \alphas\ and \water\ in Fig. \ref{fig:case1_peaks}(c) and (d) show some of the same rotation, but as mentioned above, by comparing the peak locations with the averaged plot, the overall distribution is dominated by the distribution function values with \(v_E < 0\). However, the peak locations of the distribution do still change over time. For the solar wind species, panels (a) - (d) of Fig. \ref{fig:case1_peaks} stand in stark contrast to panels (a) - (d) of Fig. \ref{fig:case2_peaks}, where the distribution changes little in time, supporting the beam-like distribution interpretation of Fig. \ref{fig:case2}. The deflection from \(-v_x\) in the \(-v_E\) direction of both solar wind species in Fig. \ref{fig:case2_peaks} is indicative of mass loaded behavior, with \proton\ being slightly more deflected than \alphas.

We do not include the low energy \water\ in Figs. \ref{fig:case1_peaks} or \ref{fig:case2_peaks} since, as previously stated, their distributions for both cases are most likely an incompletely-observed Maxwellian. For the higher energy \water, the bulk of the peak values seem to indicate a partially-filled thickened shell distribution with a wide range in energy and somewhat smaller range in pitch angle. The wide distribution in energy is likely due to differences in the pickup location, with the highest energy ions having been picked up farthest upstream of the observation point. The initial ring-beam distribution of pickup ions is generally unstable and hence is scattered in pitch angle by various types waves, which have been studied at comet 67P \citep{Szego2000, Gunell2017a, Odelstad2020}  Thus we would not expect this distribution to be narrow in phase angle, which is true for case 2 but not case 1. The bulk of the \water\ distribution for case 1 has \(v_E < 0\), indicating it is generally moving against the electric field. 

\subsection{Nongyrotropy}
Because the distribution functions for the solar wind species in case 1 are indicative of time-varying nongyrotropic effects, we also examined the diagonal and non-diagonal elements of the momentum flux tensor. As discussed above, off-diagonal elements will be comparable to diagonal elements when a plasma has a significant nongyrotropic component. Because of the small scale of the interaction region at comets, the Finite Larmor Radius effect becomes important for the momentum flux. Thus, we expect there to be non-negligible off-diagonal momentum flux components for both cases. However, the appearance of the distribution functions for both cases sets up the expectation that the nongyrotropy will be much more prominent in case 1, since the distributions are much more variable than in case 2. To examine this, we plot the ratio of the diagonal component magnitude \(\left( \sqrt{\Phi_{Pxx}^2 + \Phi_{PBB}^2 + \Phi_{PEE}^2}\right) \) to the magnitude of the off-diagonal components \(\left(\sqrt{\Phi_{PxB}^2 + \Phi_{PxE}^2 + \Phi_{PBx}^2 + \Phi_{PBE}^2 + \Phi_{PEx}^2 + \Phi_{PEB}^2}\right)\). A ratio close to 1 indicates that the magnitude of the off-diagonal elements is nearly that of the diagonal elements, and therefore the momentum flux experiences nongyrotropic effects.

\begin{figure}[ht]
    \centering
    \includegraphics[width=\hsize]{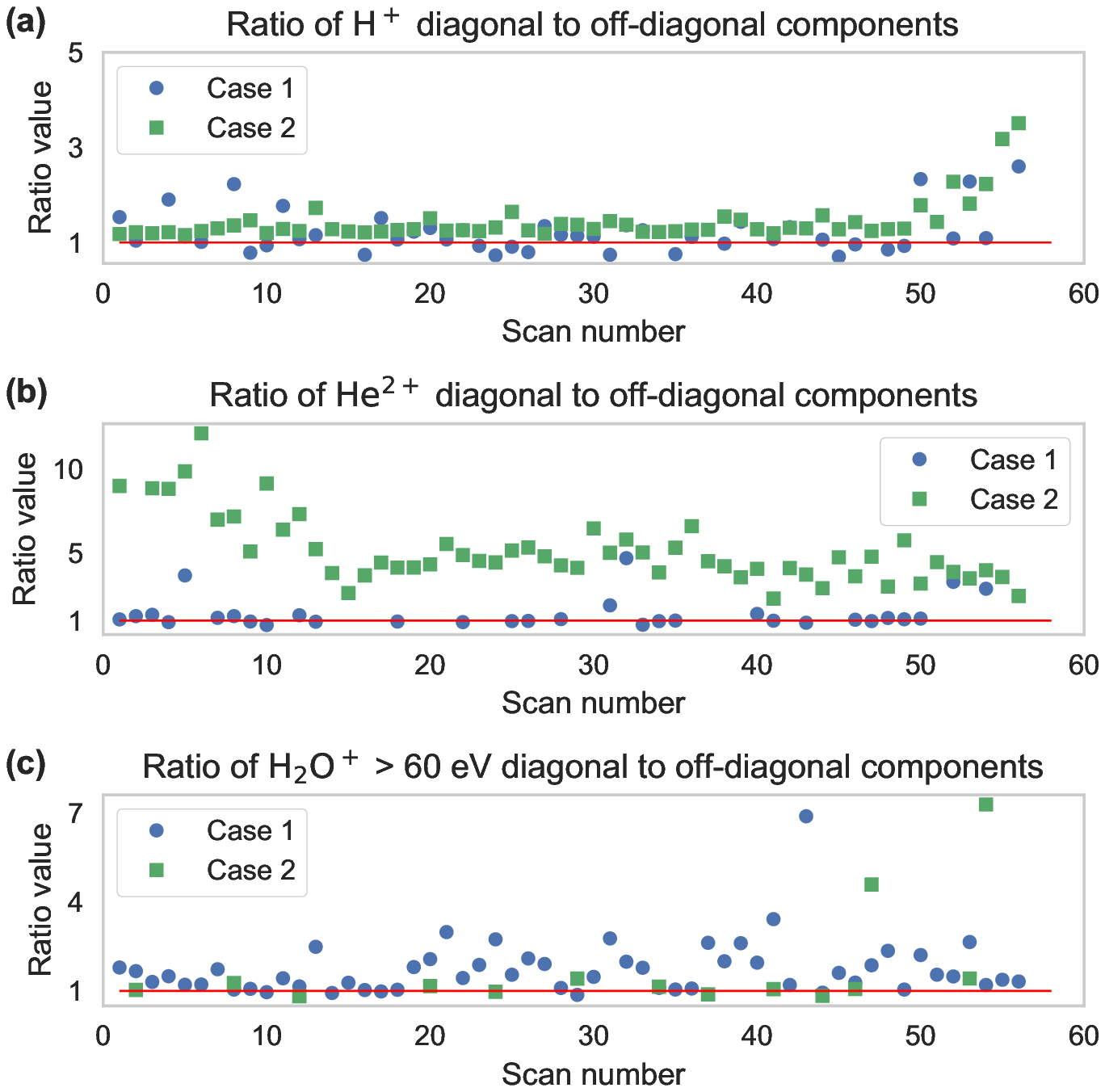}
    \caption{Ratios of the magnitude of the diagonal momentum flux components to the magnitude of the off-diagonal components for (a) \proton, (b) \alphas, and (c) \water > 60 eV plotted versus scan number. For all three, the blue circles are values for case 1 and the green squares are values for case 2. A ratio of 1 is marked by a red line on each panel.}
    \label{fig:ratios}
\end{figure}

There is a particularly large difference between the two cases for \proton\ and \alphas\ in Fig. \ref{fig:ratios}(a) and (b). For case 1, the blue circles, the ratio is occasionally < 1, indicating that the off-diagonal components are larger than the diagonal components. In contrast, for case 2 (green squares), the ratio is almost always > 1 for both solar wind species, as seen in Table \ref{tab:ratio_stats}. However, there are multiple scans in case 2 for \proton\ where the ratio is close to 1, despite the distribution being generally stationary. It is to be expected that the nongyrotropic (off-diagonal) components are present, even for the relatively undisturbed case 2, due to the scale of the comet-solar wind interaction region being small compared to the solar wind gyroradii. Thus it is likely this is simply due to the length scale. For comparison, when calculating the ion moments for the similar instrument Ion Mass Analyzer of ASPERA-3 on Mars Express, it was concluded that the off-diagonal components of the pressure tensor were so negligible that they need not be considered at all \citep{Franz2006}. Regardless of the case, the small scale of the cometary environment compared to that of, for instance, Mars, necessitates a consideration of the off-diagonal momentum flux components.

\begin{table}[ht]
    \caption{Mean values, standard deviation, and percentage of the values below 1 (n = 56) for the ratios seen in Fig. \ref{fig:ratios}.}
    \label{tab:ratio_stats}
    \centering
\begin{tabular}{lrrr}
\hline
 Species                             &   Mean &   SD &   $\% < 1$ \\
\hline
 $\mathrm{H^+}$ case 1               &   1.22 & 0.44 &      25.00 \\
 $\mathrm{H^+}$ case 2               &   1.44 & 0.43 &       0.00 \\
 \hline
 $\mathrm{He^{2+}}$ case 1           &   1.41 & 0.92 &      23.21 \\
 $\mathrm{He^{2+}}$ case 2           &   5.11 & 1.97 &       0.00 \\
 \hline
 $\mathrm{H_2O^+ > 60 \, eV}$ case 1 &   1.74 & 0.92 &       8.93 \\
 $\mathrm{H_2O^+ > 60 \, eV}$ case 2 &   1.78 & 1.78 &       7.14 \\
\hline
\end{tabular}
\end{table}
Comparing the ratios for the cases in Fig. \ref{fig:ratios} and Table \ref{tab:ratio_stats} shows a difference for the solar wind species, although the standard deviations are large because of the variability of the data. However, the ratios for \proton\ and \alphas\ are generally closer to 1 for case 1 than for case 2. Thus, both the momentum flux and velocity distribution functions indicate that nongyrotropic effects are important inside the cometopause and cometosheath. 

\section{Discussion} \label{sec:discussion}
In this section, we discuss both cases individually, then we compare the two to show that case 1 is evidence of cometosheath formation. As shown in Table \ref{tab:cases}, the magnetic field strength in case 1 is nearly twice that in case 2, indicating significant magnetic field pileup as a result of mass loading. Thus the gyroradii of the two cases differ, calculated from the magnetic field strength and the velocity component perpendicular to \(\vec{B}\):
\begin{equation}
r_g = \frac{m |v_{\perp}|}{q |B|}
\end{equation}
The mean gyroradii for each case and species, as well as the standard deviations, are shown in Table \ref{tab:gyro} for the relatively stable magnetic field periods chosen for the ion distributions. The change in gyroradius affects the ion distributions, which we discuss below. The large gyroradius for \water\ in case 2 is due to a high mean velocity, although there is considerable variability from poor statistics as evidenced by the large standard deviation.
\begin{table}[ht]
    \caption{Mean and standard deviation of the gyroradius for \proton, \alphas, and \water\ > 60 eV for case 1 14:00-17:00 and case 2 03:00-06:00.}
    \label{tab:gyro}
    \centering
\begin{tabular}{llrr}
\hline
 Species                             & |B| (nT)           &   Mean $\mathrm{r_g}$ (km) &   SD (km) \\
\hline
 $\mathrm{H^+}$ case 1               & 18.9 &  133.25 &     40.24 \\
 $\mathrm{He^{2+}}$ case 1           &      &  356.78 &     89.87 \\
 $\mathrm{H_2O^+ > 60 \, eV}$ case 1 &      &  285.87 &     85.12 \\
 \hline
 $\mathrm{H^+}$ case 2               & 6.3  &  564.59 &    142.66 \\
 $\mathrm{He^{2+}}$ case 2           &      & 1242.53 &    268.82 \\
 $\mathrm{H_2O^+ > 60 \, eV}$ case 2 &      & 3215.20 &   2981.27 \\
\hline
\end{tabular}
\end{table}

\subsection{Case 1 - January 23, 2016}

The distributions and ion moments for case 1 are different from those in previous case studies of mass loading (e.g., \citet{Behar2016,Behar2017}) in that the solar wind is not only deflected but scattered and heated, much like pickup ions. In particular, \proton\ exhibit distributions that are frequently along the magnetic and convective electric fields, unlike those studied in previous works. The \alphas\ distributions displayed in the second row of Fig. \ref{fig:case1} bear a resemblance to the high energy pickup ion distributions in the third row, albeit at higher energies, rather than \proton, whereas previous works tend to have similar distributions for the solar wind species. Notably, \alphas\ and \water\ pickup ions exhibit similar broadening in energy, as well as trajectories in the \(-v_x\) and \(-v_E\) directions. The high energy \water\ pickup ions in the third row of Fig. \ref{fig:case1} appear to have been picked up well upstream of the observation point and thus have a strongly antisunward partial shell distribution. The mean gyroradii of both species in case 1 are \(\approx 160\) km; hence, the distributions do not have space to fill a full shell distribution at the location of Rosetta, downstream of the pickup position, presumed to be thousands of km upstream from the nucleus and spacecraft. Their similar gyroradii also means that they are deflected a similar amount in the \(-v_E\) direction. Their velocity distributions, then, including movement as seen in the online video, echo each other, although they are not exactly the same.

The gyroradius of \proton\ is roughly 37\% and 47\% of the gyroradii of \alphas\ and \water\ for case 1, respectively, at \(\sim 133\) km. The threshold for nongyrotropic behavior depends on a comparison of the characteristic length scale with the circumference of a gyration \citep{Motschmann1997}, namely, \(2 \pi r_g\) or \(\approx 835\) km for \proton. At a comet, the characteristic length scale can be defined, similarly to an exobase, as the altitude region where the mean-free path of an ion is greater than its scale height. For activity levels similar to those in case 1, this altitude was generally around 100-300 km \citep{Mandt2016}. While the gyration circumference is still large enough for nongyrotropic behavior, the gyroradius is approaching the characteristic length scale and so we see partial gyration in the \proton\ distribution that is visible as a rotation of the distribution peak. Because the distribution is centered roughly at 0 km/s, the drift velocity of the \proton, \(\tfrac{\vec{E} \times \vec{B}}{B^2}\), must therefore be very low, in keeping with the presumed low bulk velocity. While we are not able to calculate \(\vec{E}\) from the available data, the increase in \(\vec{B}\) would account for a decrease in the drift velocity. If the rotation is a result of observing gyration, it may be at the \proton\ gyrofrequency, roughly 3 Hz. The ICA scan time of 192 s is then the length of nearly 100 \proton\ gyroperiods, and so we cannot confirm that the frequency of the distribution rotation seen in the online video and sequential frames in Fig. \ref{fig:seq_case1} is equal to that of the \proton\ gyrofrequency. However, the combination of a gyrofrequency that is much faster than the ICA scan time and a low drift velocity means that it is reasonable to see multiple groups of \proton\ at different phases of gyration at the position of the spacecraft. This also accounts for why there are scans with multiple phase space density peaks, which can be interpreted as observations of \proton\ in different portions of the gyromotion cycloid. 

While some of the \proton\ behavior can be attributed to observation of gyration, it does not account for all of the characteristics of the velocity distribution and moments for this case. As is visible in Fig. \ref{fig:seq_case1}, the energy and phase angle of the distribution can change significantly in the 192 s ICA scan time, indicating nongyrotropic behavior. Counter to expectations from mass loading, the \proton\ distribution is often along the \(+v_E\) direction. In a mass-loading scenario, the solar wind is initially traveling in an antisunward direction with a beam-like distribution. When it encounters the diffuse cometary atmosphere, cometary ions start to get picked up by the interplanetary magnetic field and follow the convective electric field. The ions in the coma have been picked up at various distances from the comet nucleus, with the energy gained by a pickup ion proportional to the distance from the point of ionization \citep{Nilsson2018}. Any given observation point will therefore see a range of pickup ion energies. As these pickup ions gain momentum, the solar wind must hence lose momentum and be deflected in the opposite direction, against the magnetic and convective electric field directions. However, in case 1, it is instead the \water\ pickup ions, and to a lesser extent \alphas, that are moving against the convective electric field, \(-v_x \times \vec{B}\). This is commonly seen for pickup ions when they become mass loaded by the more locally born pickup ions \citep{Nilsson2020}, which carry the majority of the momentum in this region \citep{Williamson2020}. There is additional broadening in energy of the solar wind as compared to the merely deflected distribution seen in case 2 (discussed below). The ratio of the gyrotropic to nongyrotropic momentum flux components in Fig. \ref{fig:ratios}(a) varies from scan to scan, much like the velocity distributions, and is dominated by the nongyrotropic components for half of the scans. Together, the distributions and momentum flux of \proton\ indicate significant scatter and are not consistent with changes from mass loading. 

Interestingly, the \proton\ and \water\ pickup ion distributions for this case appear to be switched compared to those observed at comet Grigg-Skjellerup by the Giotto spacecraft. At Grigg-Skjellerup, nongyrotropic distributions similar to those for \proton\ here, with a wide spread in phase angle, were seen for the \water\ pickup ions between the bow shock and cometopause, that is, in the cometosheath \citep{Coates1993,Coates1996,Motschmann1993}.  Distributions with similar dispersion in energy and angle have been observed at comet 67P as well, but were again for the \water\ pickup ions \citep{Nicolaou2017}. This may be due to the relative densities of \proton\ and \water\ for this case; here, \(n_{sw} \sim 0.07 \, n_{com}\), while for the nongyrotropic \water\ distributions at Grigg-Skjellerup, \(n_{sw} \sim 10 \, n_{com}\) \citep{Coates1997}. The difference in relative densities may also explain why \proton\ distributions here are more similar to the nongyrotropic \water\ at Grigg-Skjellerup, while the \water\ pickup ions here are deflected in the \(-v_E\) direction as with the mass-loaded solar wind, although the distribution is broader in energy. This would suggest that in this case, the \proton\ ions are enough of a minority species to serve as test particles ``picked up" by the antisunward-flowing \water, which is the reverse of typical conditions.

\subsection{Case 2 - May 10, 2016}

For this case, the total solar wind momentum flux is greater than that of the cometary pickup ions by nearly an order of magnitude. Therefore, case 2 had occurred after Rosetta crossed the cometopause. However, it is still an example of a mass-loaded solar wind. We see a solar wind that has been deflected in the \(-v_E\) direction and \water\ pickup ions that have a range of energies and are primarily traveling along the convective electric field direction, \(+v_E\), which are signatures of the early stages of mass loading. There has been little broadening in the energy of the solar wind and it still retains a relatively high velocity. This is consistent with previous observations, which show that the solar wind at comet 67P was largely deflected, rather than slowed \citep{Behar2017}.

There are some unexpected aspects that characterize case 2. Firstly, as seen in Fig. \ref{fig:case2_peaks}, the \alphas\ distribution is not completely beam-like for all scans but occasionally has two peaks in a single scan. There is a faint signal of this as well for the \proton\ in Fig. \ref{fig:case2}, but the signal is below the threshold to be considered a local maximum peak by the algorithm used to create Fig. \ref{fig:case2_peaks}. This is also visible in some frames of the scan-by-scan video. These second peaks for the \alphas\ are approximately \(180^{\circ}\) in phase angle from the primary, higher phase space density peak. Additionally, as seen in Fig. \ref{fig:ratios}, the protons have a diagonal/off-diagonal momentum flux ratio that is around 1 for the most of case 2, with a subsequent increase around 02:30. These changes occur in roughly the same time period, despite no obvious changes in the magnetic field. The densities of all the species, particularly \water, also decrease briefly near 04:00, although this time period is not included in the distribution function or momentum flux study. The unusual behavior exhibited by the plasma during this time period may indicate a transient disturbance or wave activity during this time period. However, as wave activity is not the focus of this work and these effects are not reflected in the average distribution of any of the species in Fig. \ref{fig:case2}, we do not examine this in detail. 

While the solar wind species distributions are overall similar to each other, there is more deflection in the \(+v_{B}\)and the \(-v_E\) directions for \proton\ than \alphas. This is, as in case 1, due to the smaller gyroradius of \proton. Although the magnetic pileup is more pronounced in case 1, the magnetic field strength for this case is still greater than that of the undisturbed solar wind, which is typically a couple of nT at these heliocentric distances. In addition to the deflection caused by mass loading, the \proton\ will also be affected by this increase in \(|B|\) and hence its motion will begin to have a visible gyrating component. This gyration will contribute to the \(-v_E\) deflection seen for both solar wind species. So then the \proton\ in particular, with their smaller gyroradius, will form a more visible partial ring formation, albeit one that has not undergone as much modification of the original beam-like distribution as in case 1. Similar distributions have been both observed and modeled under similar cometary activity levels for comet 67P \citep{Behar2016a,Behar2016,Behar2017}. Because of the Finite Larmor Radius effect, this distribution is still nongyrotropic, as is evident in its asymmetry in phase angle. However, unlike in case 1, the solar wind distributions here show relatively stationary nongyrotropy. Thus, the solar wind has not been disturbed from its typical narrow beam distribution.

\subsection{Cometosheath formation}

Comparing case 1 and case 2 shows that the unusual plasma characteristics in case 1 are not typical for the mass loading that occurs at a small comet. As described above, mass loading results in a deflected and, in the case of comet 67P, only slightly decelerated solar wind. Eventually, the solar wind and pickup ions will gyrate around a common center of mass, with the solar wind exhibiting a ring-beam distribution \citep{BeharThesis,Szego2000}. In case 1, we see a \proton\ distribution that does indeed appear to have been gyrating with a low drift velocity, but also displays a non-isotropic spread in phase angle, that is, a partially filled distribution, with rapid changes in energy. Consequently, the \proton\ distribution is not coupled to the convective electric field direction, with observations of the distribution function both parallel and antiparallel to \(\vec{E}\). So in case 1 \proton\ exhibits energy scattering, with the radius of the \proton\ distribution changing in sequential scans; thermalization, as indicated by the broadening of the distribution relative to that in case 2 and increased gyration; and time-varying nongyrotropy, where the distribution is highly dependent on phase angle and moves in all directions, independently of the convective electric field direction. Meanwhile, \alphas\ and \water\ pickup ions seem to be much less affected, although the \alphas\ do show some evidence of time-varying nongyrotropic distributions at times, with some scans having a faint second peak in the distribution, such as in the second and third scans in row (b) of Fig. \ref{fig:seq_case1}. 

Not only are the distributions different than expected for initial mass loading at a small comet such as 67P, but the nongyrotropic momentum flux components are larger and more inconsistent in case 1 than case 2; this runs counter to expectations based on the environment scale length and gyroradii. If the nongyrotropy was solely due to the inability of the ions to gyrate within the size of the coma, then the ratio shown in Fig. \ref{fig:ratios} should be smaller for case 2 than case 1, given the longer gyroradii and lower cometary activity in case 2. However, this is not the case for the solar wind species, as shown in Table \ref{tab:ratio_stats}. The time-varying nongyrotropic behavior of \proton\ in case 1 may be enhanced by a boundary that essentially randomizes the \proton\ phase angle, while \alphas\ and \water\ merely experience additional deflection. The nongyrotropy of the \alphas\ and \water\ in case 1 and all species in case 2 match what would be expected for the Finite Larmor Radius regime, with significant off-diagonal momentum flux components, but maintaining a velocity distribution with a relatively small spread in phase angle. The \proton\ in case 1, consequently, stands out with its frequent changes in direction and spread in phase angle and energy. 

A cometosheath, equivalent to a magnetosheath at a planet, is a transitory region downstream of a bow shock. In previous comet flybys, the cometosheath was characterized by a transition to the cometary ion density being several times that of the solar wind, a magnetic pileup region, and nongyrotropic pickup ion distributions \citep{Coates1997,Johnstone1986,Neubauer1987}. Such nongyrotropic distributions are unstable, producing waves and turbulence, which then cause the distribution to diffuse in phase angle until it reaches equilibrium \citep{Szego2000}. Thus unstable, nongyrotropic distributions are typical of a cometosheath. Additionally, a comparison of the \proton\ in case 1 and case 2 shows that, in case 1, the convective electric field is no longer the primary determinant of the \proton\ flow direction, as it is in the mass loading at a small comet scenario. In case 1, gyration is now primarily responsible for the \proton\ velocity distribution function. The transition of the solar wind from deflection against the convective electric field seen in case 2 to gyration is a hallmark of magnetosheath formation, as gyration and wave-particle interactions lead to the heating seen in a magnetosheath \citep{Nilsson2012,Nilsson2017}. 

While such time-varying nongyrotropic distributions and independence from the convective electric field are not strongly evident for species other than \proton, despite \water\ pickup ions being the species that exhibited such behavior at comet Grigg-Skjellerup, this is not unprecedented for a magnetosheath. The possibility of the \proton\ distribution in case 1 being caused by a boundary that the other species are capable of passing through due to larger gyroradii is supported by studies of solar wind precipitation at Mars. Precipitation of \proton\ and \alphas\ was seen at lower altitudes in the Martian magnetosheath when the magnetosheath distance was small compared to the solar wind species gyroradius. Thus the higher energy particles were able to pass through the magnetosheath barrier \citep{Dieval2012,Stenberg2011}. Further studies have shown that an increase in solar wind dynamic pressure led to a decrease in solar wind precipitation, as the magnetic pileup region widened with respect to the \proton\ and \alphas\ gyroradii and hence this gyration around was no longer possible \citep{Dieval2013}. 

The difference between \proton\ and \alphas\ in case 1 indicates that a scenario similar to the precipitation at the Martian magnetosheath is likely occurring for higher activity levels at comet 67P. The \alphas (as well as \water\ pickup ions) with their longer gyroradius are able to ignore more of the effects of the magnetic barrier formed by the solar wind magnetic field pileup in a cometosheath. Meanwhile \proton\ begin to gyrate within the cometosheath and are subject to the scattering and heating from wave-particle interactions that occur with unstable velocity distributions. The narrow width of the cometosheath boundary imposed by the difference between the \proton\ gyroradius and the \alphas\ gyroradius suggests a sudden transition, and is possibly further evidence that comet 67P did develop a weak bow shock. The narrow width of the cometosheath compared to broad cometosheaths seen at Halley and Grigg-Skjellerup \citep[e.g.,][]{Huddleston1993,Neugebauer1990a} is due to the overall lower activity of comet 67P, such that the interaction regions are generally smaller than those seen by previous cometary missions. Additionally, the narrow cometosheath suggests that case 1 is on the edge of the activity levels required for a shock, and subsequently a cometosheath, to form. The case for the formation of a narrow shock and cometosheath is also supported by the broadening of both the \proton\ and \alphas\ distributions in energy, which has been previously argued as evidence of thermalization of the solar wind due to a bow shock \citep{Goetz2021}. However, given the high solar wind velocities and long ICA scan times, such that the solar wind can travel many thousands of kms within a single scan, prevent us from being able to resolve such a narrow boundary in the ICA data. The evidence of only weak thermalization, as well as the transition to gyromotion for primarily \proton, indicate that case 1 is on the lower limit of the activity levels required for shock and cometosheath formation. Therefore, case 1 is a boundary case showing that it was indeed possible for a cometosheath, downstream of a bow shock that was capable of modifying the solar wind, to be observed in the ICA data, even if the shock itself cannot be resolved.

\section{Conclusions} \label{sec:conclusions}

In this work, we compared two data sets from the Rosetta mission to comet 67P/Churyumov-Gerasimenko from January 23, 2016 and May 10, 2016. We compared the velocity distribution functions and the momentum flux to understand the differences between two parts of the comet-solar wind interaction region.

The first case was during the time period when ICA saw a plasma that was dominated in momentum flux by the cometary ions and is an example of medium-high cometary activity, with gas production rates approximately one order of magnitude lower than comet Grigg-Skjellerup. ICA observations show time-varying \proton\ distributions that have been scattered in phase angle and energy, and are often directed along the convective electric field. Meanwhile, \alphas\ and \water\ pickup ions, while also showing evidence of heating, show less rotation and similar spread in phase angle, with the flow generally in the \(-v_x, -v_E\) direction. We observe \proton\ in different phases of gyration, accounting for the rotation of the \proton\ distribution. We also find that all three species show nongyrotropic momentum flux, but that the nongyrotropy is more prevalent for \proton. 

The second case, for lower cometary activity, shows an example of mass loading when the momentum flux is dominated by the solar wind. The \water\ pickup ions are primarily moving along the convective electric field, with a broad spread in energy due to observations of ions picked up at different distances from the spacecraft. The solar wind species are subsequently deflected in the \(-v_E\) direction with little change in velocity. The momentum flux does still contain nongyrotropic elements, particularly for the \water\ pickup ions, but they contribute less to the solar wind momentum flux than in case 1.

Comparing the first case to the second shows that mass loading alone cannot account for the changes in the ion distributions and momentum flux seen in case 1. Firstly, while a broadened distribution in energy is expected for pickup ions, we also see a broadened distribution for \alphas\ and \proton. Secondly, mass loading tends to deflect the solar wind distribution in the \(-v_E\) direction, which is not seen for \proton\ in case 1. Instead, the \proton\ distribution varies widely in phase angle, energy, and time, unlike that seen in case 2. We also see more strongly nongyrotropic momentum fluxes for the solar wind species in case 1 than case 2, even though a greater extent of nongyrotropy would be expected for case 2. We therefore interpret case 1 to indicate the formation of a cometosheath with a width on the order of the \proton\ gyroradius. Then \proton\ would be able to make at least a partial gyration with a low drift velocity within this narrow cometosheath, while \alphas\ and \water\ pickup ions are less affected due to their larger gyroradii. The cometosheath-like distributions indicate the development of a bow shock upstream of the observations that causes the scatter and heating of the ions, although it is too narrow to be resolved in ICA data.

\begin{acknowledgements}
The Rosetta RPC-ICA and RPC-MAG data are publicly available through the Planetary Science Archive of ESA at https://archives.esac.esa.int/psa/. The work in this study was funded by the Swedish Research Council contract 2015-04187 and Swedish National Space Agency grant 132/19. Color maps used courtesy of CMasher \citep{CMR_colors}. We thank the members of the ICA and LAP team meetings as well as the IRF SSPT group for their discussions on this topic.
\end{acknowledgements}

\bibliographystyle{aa} 
\bibliography{references} 
\end{document}